\documentclass[12pt]{article}
\usepackage{amsmath,amssymb,amsfonts,amsthm,mathrsfs,verbatim,accents}
\usepackage{graphicx}
\usepackage{hyperref}
\hypersetup{colorlinks,linkcolor={blue},citecolor={magenta},urlcolor={magenta}} 
\usepackage{cite}
\usepackage[english]{babel}
\usepackage[utf8]{inputenc}
\usepackage{listings}
\usepackage[T1]{fontenc}
\usepackage{lmodern}
\usepackage{caption}
\usepackage{floatrow}
\usepackage{subcaption}
\usepackage{mathrsfs}
\usepackage{enumerate}
\usepackage[normalem]{ulem}
\usepackage{color}
\usepackage{array}
\usepackage{xcolor}

\begin{document}

\title{\hfill\mbox{\small}\\[-1mm]
\hfill~\\[0mm]
       \textbf{Connecting Tribimaximal and Bitrimaximal Mixings}        }
\date{}
\author{\\[1mm] 
	Carlos Alvarado$^{1,2\,}$\footnote{E-mail: {\tt calvarado@vassar.edu}}~,
	Janelly Bautista$^{3,4\,}$\footnote{E-mail: {\tt bauti069@umn.edu }}~,~and Alexander J.~Stuart$^{2,3\,}$\footnote{E-mail: {\tt astuart@ucol.mx}}\\
	\\[1mm]
	\textit{\small $^1 $Physics and Astronomy Department, Vassar College,}\\
	\textit{\small Box 745, Poughkeepsie, NY 12604, U.S.A.}\\[3mm]	
	\textit{\small $^2$Dual CP Institute of High Energy Physics,}\\
	\textit{\small C.P.~28045, Colima, M\'exico}\\[3mm]
	\textit{\small $^3 $Facultad de Ciencias-CUICBAS, Universidad de Colima,}\\
	\textit{\small C.P.~28045, Colima, M\'exico}\\[3mm]
	\textit{\small $^4 $School of Physics and Astronomy,  University of Minnesota,}\\
	\textit{\small Minneapolis, MN 55455, U.S.A.}\\[3mm]
}
\maketitle

\vspace{0.5cm}

\begin{abstract}
\noindent
In this paper, we study the connection between the tribimaximal and bitrimaximal mixing patterns.  In doing so, we are forced to work in a non-diagonal charged lepton basis.  This leads to several relations that must hold between the lepton mixing angles.  After a short discussion, we analyze the underlying flavor symmetry responsible for this prediction.  Finally, we add CP violation to bitrimaximal mixing and study its effect on the flavor symmetry group.

\end{abstract}
\thispagestyle{empty}
\vfill
\newpage
\setcounter{page}{1}

%%%%%%%%%%%%%%%%
\newpage
%%%%%%%%%%%%%%%%

%%%%%%%%%%%%%%%%%%%%%%%%%%%%%%%%%%%%%%%%%%%%%%%%%%%%%%%

\section{Introduction\label{sec:intro}}
The confirmation of a sizable reactor mixing angle by the Daya Bay \cite{dayabay}, RENO\cite{reno}, and Double Chooz\cite{doublechooz} Collaborations has further added intrigue to the already fascinating flavor puzzle.  The Cabibbo-sized value of this angle has now become the harbinger of a possible measurement of CP violation in the lepton sector.  Actually, current refined measurements \cite{T2KdeltaCP,NOvadeltaCP} suggest not only a non-zero \textit{Dirac-type} CP-violating phase, $\delta$,  but a phase whose magnitude leans towards $|\delta|=\pi/2$. However, a more conservative view based on global fits \cite{nufitpub,tortola,lisi,nufitweb} still leaves $\delta$ rather undetermined  at $3\sigma$.  

The arguably most enticing approach to understand the parameters which populate the Maki-Nakagawa-Sakata-Pontecorvo (MNSP) leptonic mixing matrix $U_{\text{MNSP}}$ \cite{mnsp, pdg} is one based on discrete flavor symmetries \cite{kingluhnreview, flavorring, feruglioromaninoreview, asymmetric, fonsecagrimus, hernsmirn, delta6nsq, Z2CP1}.   Usually, a larger discrete flavor symmetry group is spontaneously broken to a residual subgroup  which can generate mixing patterns like TriBiMaximal (TBM) mixing\cite{tbm}, Golden Ratio mixing (GR1) \cite{GR1, A5GR1}, BiMaximal (BM) \cite{bimaximal, S4bimax}, and BiTriMaximal (BTM) mixing \cite{btmmixorig, BTmixing, delta96pres}.
This has led to previous studies of the interplay between mixing patterns, residual lepton symmetries, and CP-violating phases \cite{residualreviewsChenValle}. This work will focus primarily on TBM mixing and BTM mixing. These two mixing patterns' names originate from what was originally called TriMaximal (TM) mixing\cite{trimaxorig1,trimaxorig2} which has $|(U_{\text{MNSP}})_{ij}|^2=1/3$, where $i,j=1,2,3$.  Thus, it has a threefold maximal mixing.  From this followed TBM mixing in which the MNSP matrix has a trimaximal middle column and a bimaximal third column.  After this followed BTM mixing which has a trimaximal middle column and trimaximal middle row (named in Ref.~\cite{delta96pres}).  Notice that both TBM and BTM mixing patterns have a trimaximal middle column.  This similarity will be exploited to relate these patterns, and to do so, we restrict ourselves to BTM mixing by regarding it as a leading-order prediction for the current neutrino data. Then, model-dependent corrections to the mixing angles and CP-violating phase can be used to bring the mixing parameters closer to measurements. Yet, such model-dependent corrections fall outside the scope of this work.

The remainder of this work is as follows.  In Section \ref{sec:framework}, we review a particular framework that helps to understand residual lepton symmetries, which was first developed in Ref.~\cite{bottomup}.  We then use this framework to derive relationships which must be satisfied among the mixing angles of the MNSP matrix in order to obtain a trimaximal middle column.  We next relate TBM mixing to BTM mixing by utilizing the definition that  $U_{\text{MNSP}}=U_e^{\dagger}U_{\nu}$, where $U_e$ is the unitary matrix that diagonalizes the charged lepton mass matrix, and $U_{\nu}$ is the unitary matrix that diagonalizes the neutrino mass matrix. This is done by assuming that $U_{\nu}$ is TBM mixing, $U_{\text{MNSP}}$ is BTM mixing, and then deriving the resulting $U_e$ which connects the two.  Some implications of this non-diagonal $U_e$ on the charged lepton mass matrix  are then explored, and the group-theoretical structure of the non-diagonal $U_e$ and TBM $U_{\nu}$ is derived.  In Section \ref{sec:btmphase}, we impose a non-trivial CP-violating phase in the BTM mixing matrix and repeat the logic of Section \ref{sec:framework} to reveal the discrete symmetry group associated with the CP-violating BTM mixing pattern and comment on the result.  Finally, in Section \ref{sec:conclusions} we conclude.

%%%%%%%%%%%%%%%%%%%%%%%%%%%%%%%%%%%%%%%%%%%%%%%%%

\section{\label{sec:framework}  Framework and Motivation}

%%%%%%%%%%%%%%%%%%%%%%%%%%%%%%%%%%%%%%%%%%%%%%%%%%%%%%%%%%%%%%%%%%%%%%%%%%%

%%%%%%%%%%%%%%%%%%%%%%%%%%%%%%%%%%%%%%%%%%%%%%%%%%%%%%%%%%%%%%%%%%%%%%%%%%%

\subsection{\label{subsec:z2xz2}  The Neutrino Sector and $\boldsymbol{Z_2\times Z_2}$}

%%%%%%%%%%%%%%%%%%%%%%%%%%%%%%%%%%%%%%%%%%%%%%%%%%%%%%%%%%%%%%%%%%%%%%%%%%%
We begin this discussion with an overview of the $Z_2\times Z_2$ symmetry in neutrino mixing.  By assuming that neutrinos are Majorana fermions (see Refs.~\cite{kingluhnreview} and \cite{feruglioromaninoreview} for reviews), the neutrino mass matrix $M_{\nu}$ is a complex symmetric matrix, i.e., $M_{\nu}=M_{\nu}^T$.  Thus, it can be diagonalized by a unitary matrix, $U_{\nu}$, such that
\begin{equation}\label{eq:Mnudiag}
	U_{\nu}^T M_{\nu}U_{\nu}=M_{\nu}^\text{Diag}=\text{Diag}(|m_1| e^{i\alpha_1}, |m_2| e^{i\alpha_2}, |m_3| e^{i\alpha_3})=\text{Diag}(m_1,m_2,m_3),
\end{equation}
where $\alpha_{1,2,3}$ contribute to the two physical Majorana phases\cite{majorana1,majorana2}.  The utility of keeping the neutrino masses complex has been demonstrated in Refs.~\cite{bottomup, bottomupext}.  Notice that the neutrino masses can be made real and positive with the replacement of $U_{\nu}\rightarrow U_{\nu}P_{\text{Maj}}$, where $P_{\text{Maj}}=\text{Diag}(e^{i\alpha_1/2},e^{i\alpha_2/2},e^{i\alpha_3/2})$.

In contrast to the neutrinos, the charged leptons cannot be assumed to be Majorana fermions.  Thus, the diagonalization of the charged lepton mass matrix $m_e$ can be achieved by defining  $M_e=m_e m_e^{\dagger}$ and then diagonalizing this Hermitian matrix by a unitary matrix $U_e$ such that
\begin{equation}\label{eq:Mediag}
	U_{e}^{\dagger} M_{e}U_{e}=M_{e}^\text{Diag}=\text{Diag}(|m_e|^2,|m_{\mu}|^2,|m_{\tau}|^2).
\end{equation}
In what follows, $U_{\nu}$ will be referred to as the neutrino mixing matrix and $U_e$ as the charged lepton mixing matrix.  However, recall that it is only possible to measure the overlap of these two mixing matrices, i.e., the MNSP lepton mixing matrix $U_{\text{MNSP}}=U_e^{\dagger}U_{\nu}$ \cite{mnsp,pdg}. Expressing this in the standard PDG parameterization\cite{pdg}, but rephasing it by the charged lepton phase matrix $P_e$ \cite{bottomup} yields
\begin{equation}\label{eq:mnspmat}
\begin{aligned}
U_{\text{MNSP}}=&P_e U_{\text{MNSP}}^{\text{PDG}}=P_e R_x(\theta_{23})R_y(\theta_{13},\delta)R_z(\theta_{12})\\
=&\left(
\begin{array}{ccc}
c_{12} c_{13} & c_{13} s_{12} & e^{-i \delta } s_{13} \\
-c_{23} s_{12}-c_{12} e^{i \delta } s_{13} s_{23} & c_{12} c_{23}-e^{i \delta } s_{12} s_{13} s_{23} & c_{13} s_{23} \\
-s_{12} s_{23}+c_{12} c_{23} e^{i \delta } s_{13} & c_{12} s_{23}+c_{23} e^{i \delta } s_{12} s_{13} & -c_{13} c_{23} \\
		\end{array}
		\right),
\end{aligned}
\end{equation}
where
\begin{eqnarray}
	\label{euler}
	\begin{array}{cc}
		R_{x}(\theta_{23})=\left(
		\begin{array}{ccc}
			1 & 0 & 0 \\
			0 & c_{23} & s_{23} \\
			0 & -s_{23} & c_{23}
		\end{array}
		\right),&
		R_{y}(\theta_{13},\delta)=\left(
		\begin{array}{ccc}
			c_{13} & 0 & s_{13}e^{-i \delta} \\
			0 & 1 & 0 \\
			-s_{13}e^{i \delta} & 0 & c_{13}
		\end{array}
		\right), \\&\\
		R_{z}(\theta_{12})=\left(
		\begin{array}{ccc}
			c_{12} & s_{12} & 0 \\
			-s_{12} & c_{12} & 0 \\
			0 & 0 & 1
		\end{array}
		\right),&
		P_e=\left(
		\begin{array}{ccc}
			1 & 0 & 0 \\
			0 & 1 & 0 \\
			0 & 0 & -1
		\end{array}
		\right),
	\end{array}
\end{eqnarray}
in which  $s_{ij}=\sin(\theta_{ij})$ and $c_{ij}=\cos(\theta_{ij})$.

%%%%%%%%%%%%%%%%%%%%%%%%%%%%%%%%%%%%%%%%%%%%%%%%%%%%%%%%%%%%%%%%%%%%%%%%%%%

\subsection{\label{subsec:tbmtobtm}  From Tribimaximal to Bitrimaximal}

%%%%%%%%%%%%%%%%%%%%%%%%%%%%%%%%%%%%%%%%%%%%%%%%%%%%%%%%%%%%%%%%%%%%%%%%%%%
Before the exclusion of a non-zero reactor mixing angle, the tribimaximal pattern\cite{tbm}, i.e., 
\begin{align}\label{eq:tbmmnsp}
	U^{\text{TBM}}=
	\left(
	\begin{array}{ccc}
		\sqrt{\frac{2}{3}}  & \frac{1}{\sqrt{3}} & 0 \\
		-\frac{1}{\sqrt{6}} & \frac{1}{\sqrt{3}} & \frac{1}{\sqrt{2}} \\
		-\frac{1}{\sqrt{6}} & \frac{1}{\sqrt{3}} & -\frac{1}{\sqrt{2}} \\
	\end{array}
	\right),
\end{align}
 was an attractive leading order mixing pattern
 %\footnote{Generally, Cabibbo-sized corrections to the initial angle values can be expected\cite{cabibbohaze}.}
 for the MNSP matrix because it predicts
\begin{equation}
	\theta_{12}^{\text{TBM}}=\cos^{-1}{(\sqrt{2/3})}\approx 35.24^{\circ},~~~\theta_{13}^{\text{TBM}}=0^{\circ},~~~\theta_{23}^{\text{TBM}}=\pi/4=45^\circ{}. \label{eq:tbmmixing}
\end{equation}
Thus, it was not unreasonable to construct a flavor model which predicted tribimaximal mixing at leading order and had model-dependent corrections to provide the necessary deviations. The implementation of model-dependent corrections which bring the mixing angles to the NuFIT 5.3 allowed ranges departs from the objective of our work and will not be pursued here. However, it is known that some strategies to do so exist, for example, charged lepton corrections, renormalization group running, and canonical (kinetic) normalization effects \cite{correctionsCollection,runningAntusch}. All of these are expected to be Cabibbo-sized in realistic models of neutrino mass \cite{cabibbohaze}. Many people showed that this pattern naturally emerged from groups like $S_4$ and $A_4$.\footnote{For technical reasons related to $Z_2\times Z_2$ subgroups \cite{delta6nsq}, we restrict ourselves to the 24 element group $S_4$.}   Yet, after the reactor mixing angle was empirically shown to be inconsistent with zero, many model builders looked for other starting points which had sizable predictions for the reactor mixing angle\cite{btmmixorig, delta6nsq}.  For the remainder of this work, we will specifically focus on bitrimaximal mixing\cite{btmmixorig, BTmixing},
\begin{align}\label{eq:btmmnsp}
	U^{\text{BTM}}=\frac{1}{\sqrt{3}}
	\left(
	\begin{array}{ccc}
		\frac{1}{2} \left(1+\sqrt{3}\right) & 1 & \frac{1}{2} \left(\sqrt{3}-1\right) \\
		-1 & 1 & 1 \\
		\frac{1}{2} \left(1-\sqrt{3}\right) & 1 & \frac{1}{2} \left(-1-\sqrt{3}\right) \\
	\end{array}
	\right),
\end{align}
where
\begin{equation}
	\theta_{12}^{\text{BTM}}=\theta_{23}^{\text{BTM}}=\tan^{-1}\left(\sqrt{3}-1\right)\approx 36.21^{\circ},~~~\theta_{13}^{\text{BTM}}=\sin ^{-1}\left(\frac{1}{6} \left(3-\sqrt{3}\right)\right)\approx 12.20^{\circ}. \label{eq:btmmixing}
\end{equation}
This pattern was shown to naturally originate from the discrete group $\Delta(96)$\cite{btmmixorig,BTmixing,delta96pres}.  Notice that this group is slightly more complicated than the $S_4$ group associated with tribimaximal mixing.  However, this slight increase in the size of the group allows for a non-zero leading order prediction for the reactor mixing angle of approximately $12.2^{\circ}$, see Eq.~\eqref{eq:btmmixing}.  Yet, as different as the two  mixing patterns are, they also have something in common, i.e., the trimaximal middle column\cite{trimaxorig1,trimaxorig2}.  It is well known\cite{kingluhnreview, delta6nsq} that this column can originate from the preservation of the $Z_2\times Z_2$ group element
\begin{align}
\frac{1}{3}\left(
	\begin{array}{ccc}
		-1 & 2 & 2 \\
		2 & -1 & 2 \\
		2 & 2 & -1 \\
	\end{array}
	\right).  \label{eq:canonicalS}
\end{align}

\begin{figure}[h!]
	\centering
	\includegraphics[scale=0.85]{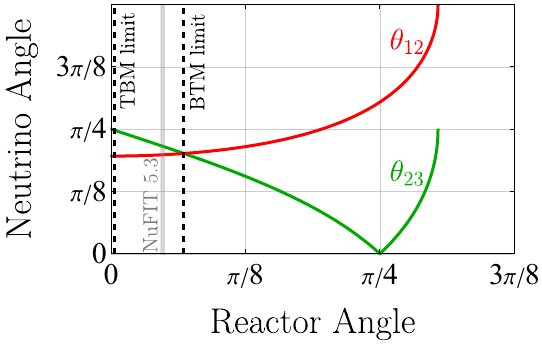}
	\caption{The parametric conditions in Eq.~\eqref{eq:trimaxconds} revealing the relationships among the mixing angles which most hold to yield a trimaximal middle column.  The dashed lines indicate the tribimaximal (TBM) and bitrimaximal (BTM) limits. The gray band shows the $3\sigma$ reactor angle constraint from NuFIT 5.3 \cite{nufitpub,nufitweb}.}
	\label{fig:mxplot}
\end{figure}

In order to understand the trimaximal mixing pattern, we consider the $Z_2\times Z_2 $ symmetry elements $G_i$ where $i=1,2,3$, as defined in Ref.~\cite{bottomup}.  Specifically, we use the angle-dependent form for the $G_2$ element, i.e., 
\begin{equation}
	\begin{aligned}
	\label{eq:G2}
		G_2=\left(
		\begin{matrix}
			\left(G_2\right)_{11} & \left(G_2\right)_{12} &\left(G_2\right)_{13} \\
			\left(G_2\right)_{12}^* &\left(G_{2}\right)_{22} &  \left(G_{2}\right)_{23} \\
			\left(G_2\right)_{13}^*&  \left(G_{2}\right)_{23}^*& \left(G_{2}\right)_{33}
		\end{matrix}
		\right),
		\end{aligned}
\end{equation}
where
\begin{equation}\label{eq:G2elements}
	\begin{aligned}
		\left(G_2\right)_{11}&= -c'_{12} c_{13}^2-s_{13}^2,\\
		\left(G_2\right)_{12}&= 2 c_{13} s_{12} \left(c_{12} c_{23}-e^{-i \delta } s_{12} s_{13} s_{23}\right), \\
		\left(G_2\right)_{13}&= 2 c_{13} s_{12} \left(e^{-i \delta } c_{23} s_{12} s_{13}+c_{12} s_{23}\right),\\
		\left(G_2\right)_{22}&= c'_{12} c_{23}^2-s_{23}^2 \left(c_{13}^2+s_{13}^2 c'_{12}\right)-\cos (\delta ) s_{13} s'_{12} s'_{23}, \\
		\left(G_2\right)_{23}&=e^{-i \delta } s_{13} s'_{12} c_{23}^2+\frac{1}{4} s'_{23} \left(2 c_{13}^2-c'_{12} \left(c'_{13}-3\right)\right) - e^{i \delta }  s'_{12} s_{13} s_{23}^2,
		\\		\left(G_2\right)_{33}&=-c_{23}^2\left(c_{13}^2+s_{13}^2 c'_{12}\right) +s_{23}^2 c'_{12}+\cos (\delta ) s_{13} s'_{12} s'_{23}, 
	\end{aligned}
\end{equation}
with $s_{ij}=\sin(\theta_{ij})$, $c_{ij}=\cos(\theta_{ij})$, $s'_{ij}=\sin(2\theta_{ij})$, and $c'_{ij}=\cos(2\theta_{ij})$. Then, demanding the entries of $G_{2}$ to take the forms of those in the $S$ in Eq.~\eqref{eq:canonicalS} preserves the trimaximality in the middle column of the MNSP matrix through the following relationships among the mixing parameters:

\begin{equation}\label{eq:trimaxconds}
\begin{aligned}
	\cos(\theta_{23}) &= \frac{1}{2} \sec (\theta_{13}) \left(\sin (\theta_{13})+\sqrt{3 \cos ^2(\theta_{13})-1}\right),  \\
	\cos(\theta_{12}) &= \frac{\sec (\theta_{13})\sqrt{3 \cos (2 \theta_{13})+1} }{\sqrt{6}},\\
	\delta &= 0.
\end{aligned}
\end{equation}
The conditions above can now be plotted to reveal the interplay between the solar and atmospheric mixing angles as a function of the reactor mixing angle.  The results of this exercise can be found in Fig.~\ref{fig:mxplot}, which shows the NuFIT 5.3 $3\sigma$ band for the reactor angle \cite{nufitpub,nufitweb}.
 It is important to note that the curves for the $\theta_{12}$ (solar) and $\theta_{23}$ (atmospheric) in Fig.~\ref{fig:mxplot} do not approach an asymptote. Instead, the radicands of the relations in Eq.~\eqref{eq:trimaxconds} become negative.  A quick calculation shows that $\theta_{12}$ and $\theta_{23}$ become complex when $\theta_{13}>\cos^{-1}(1/\sqrt{3})=1/2\cos^{-1}(-1/3)\approx 0.955$.  Lastly from Fig.~\ref{fig:mxplot} it is straightforward to deduce that there exists a continuous transformation from the tribimaximal to the  bitrimaximal mixing angle values.  In the next section, a physically motivated transformation will be revealed which naturally maps from tribimaximal mixing to bitrimaximal mixing.

%%%%%%%%%%%%%%%%%%%%%%%%%%%%%%%%%%%%%%%%%%%%%%%%%%%%%%%%%%%%%%%%%%%%%%%%%%%%%%%%%%%%%%%%%%%%%%%%%%%%%%%%%%%%%%%%%%%%%%%%%%%%%%%%%%%%%%%%%%%%%%%%%%%%%%

\subsection{The $\boldsymbol{U^{\text{TBM}}}$ and $\boldsymbol{U^{\text{BTM}}}$ Connection\label{subsec:connection}}

%%%%%%%%%%%%%%%%%%%%%%%%%%%%%%%%%%%%%%%%%%%%%%%%%%%%%%%%%%%%%%%%%%%%%%%%%%%%%%%%%%%%%%%%%%%%%%%%%%%%%%%%%%%%%%%%%%%%%%%%%%%%%%%%%%%%%%%%%%%%%%%%%%%%%%
The previous discussion hinted at a possible connection between the  tribimaximal and bitrimaximal mixing angle values when considering the conditions on the mixing angles that are needed to preserve a trimaximal middle column, see Eq.~\eqref{eq:trimaxconds}.  For the remainder of this section, we will further elaborate on their relationship.  To begin this discussion, assume that tribimaximal mixing solely originates in the neutrino sector such that $U_{\nu}=U^{\text{TBM}}$, cf.~Eq.~\eqref{eq:tbmmnsp}.  Furthermore, assume that the MNSP lepton mixing matrix has the bitrimaximal form given in Eq.~\eqref{eq:btmmnsp}.  Then, it is possible to relate the bitrimaximal MNSP matrix to the tribimaximal neutrino matrix through the charged lepton mixing matrix, $U_e$, such that
\begin{equation}
U_{\text{MNSP}}=U_e^{\dagger}U_{\nu}=U_e^{\dagger}U^{\text{TBM}}\equiv U^{\text{BTM}}. \label{eq:tbmtobtm}
\end{equation}
A straightforward calculation reveals that 
\begin{align}
U_{e}= U^{\text{TBM}}(U^{\text{BTM}})^{\dag}=\frac{1}{6}\left(
\begin{array}{ccc}
 2+\sqrt{2}+\sqrt{6} & 2-2 \sqrt{2} & 2+\sqrt{2}-\sqrt{6} \\
 2+\sqrt{2}-\sqrt{6} & 2+\sqrt{2}+\sqrt{6} & 2-2 \sqrt{2} \\
 2-2 \sqrt{2} & 2+\sqrt{2}-\sqrt{6} & 2+\sqrt{2}+\sqrt{6} \\
\end{array}
\right). \label{eq:Utbtobt}
\end{align}
Notice that this charged lepton mixing matrix has the rather peculiar property that in addition to being unitary, i.e., $U_eU_e^{\dagger}=U_e^{\dagger}U_e=1$,  the sum of all of the entries of each row as well as the sum of all of the entries of each column is non-trivially equal to unity, i.e.,
\begin{equation}
\sum_{i=1}^{3}(U_{e})_{ik}=1,~k=1,2,3~~\text{and}~~~\sum_{j=1}^{3}(U_{e})_{kj}=1,~k=1,2,3. \label{eq:sumrowcolumm}
\end{equation}
Now that the matrix which diagonalizes the charged lepton mass matrix $M_e$ has been deduced, we next look at the implications it has on said matrix.

\subsection{Implications for Charged Lepton Mass Matrix\label{subsec:implicationsCharged}}

In order to reveal the corresponding (Hermitian) charged lepton mass matrix, $M_e\equiv m_em_e^{\dagger}$, that is diagonalized by $U_{e}$ (see Eq.~\eqref{eq:Utbtobt}), we set up the relation
\begin{equation}\label{eq:Mediagonalize}
M_e^{\text{Diag}}=U_e^{\dagger}M_eU_e=\text{Diag}\{|m_e|^2,|m_{\mu}|^2,|m_{\tau}|^2\}
\end{equation}
and solve for $M_{e}$. The resulting $M_{e}$ does not have an illuminating form, so we parameterize it as
\begin{align}
&M_e= \notag \\ 
&\left( 
\begin{array}{ccc}
  a^2 \left| m_e\right| ^2+b^2 \left| m_{\tau }\right| ^2+c^2 \left| m_{\mu }\right| ^2 & a b \left| m_e\right| ^2+a c \left| m_{\mu }\right| ^2+b c \left| m_{\tau }\right| ^2 & a b \left| m_{\tau }\right| ^2+a c \left| m_e\right| ^2+b c \left| m_{\mu }\right| ^2 \\
  a b \left| m_e\right| ^2+a c \left| m_{\mu }\right| ^2+b c \left| m_{\tau }\right| ^2 & a^2 \left| m_{\mu }\right| ^2+b^2 \left| m_e\right| ^2+c^2 \left| m_{\tau }\right| ^2 & a b \left| m_{\mu }\right| ^2+a c \left| m_{\tau }\right| ^2+b c \left| m_e\right| ^2 \\
  a b \left| m_{\tau }\right| ^2+a c \left| m_e\right| ^2+b c \left| m_{\mu }\right| ^2 & a b \left| m_{\mu }\right| ^2+a c \left| m_{\tau }\right| ^2+b c \left| m_e\right| ^2 & a^2 \left| m_{\tau }\right| ^2+b^2 \left| m_{\mu }\right| ^2+c^2 \left| m_e\right| ^2 \\
\end{array}\label{eq:undiagMe}
\right),
\end{align}
where 
\begin{equation}
  \begin{aligned}
a&\equiv \frac{1}{6}(2+\sqrt{2}+\sqrt{6})=0.977\approx \lambda^0,\\
b&\equiv \frac{1}{6}(2+\sqrt{2}-\sqrt{6})=0.161\approx\lambda,\\
c&\equiv \frac{1}{6} (2-2\sqrt{2})=-0.138\approx-\lambda,
\end{aligned} \label{eq:abcdef}
\end{equation}
are hierarchical in terms of powers of the Cabibbo angle, $\lambda\approx0.2265$.  Notice that with these definitions for $a$, $b$, and $c$, Eq.~\eqref{eq:Utbtobt} becomes
\begin{align}
	U_{e}= \left(
	\begin{array}{ccc}
		a & c & b \\
		b & a & c \\
		c & b & a \\
	\end{array}
	\right). \label{eq:Utbtobtabc}
\end{align}

 Observe that the $M_{e}$ from Eq.~\eqref{eq:undiagMe}  contains only $6-1=5$ degrees of freedom because of the constraint that $a+b+c=1$ from Eq.~(\ref{eq:Utbtobt}). However, this matrix mass is cumbersome, and one could argue that the entries of this matrix should be dominated by the (squared) tau  lepton mass, $m_{\tau}^{2}$, so that the (squared) muon and electron masses
\begin{equation} \label{eq:cabbibosup}
m_{\mu}^{2}\approx (\lambda^2 m_{\tau})^{2},~~~m_e^{2}\approx(\lambda^5 m_{\tau})^{2},
\end{equation}
can be thought of as perturbations from the limit in which they vanish. In fact, in this limit $M_{e}$ acquires a rather simple form, i.e., 
\begin{equation}
M_e=\left| m_{\tau }\right|^2
\left(
\begin{array}{ccc}
  b^2 & b c & a b\\
  b c  & c^2  & a c  \\
  a b & a c  & a^2 \\
\end{array}
\right)+\left| m_{\tau }\right|^2\mathcal{O}(\lambda^4). \label{eq:undiagMelim}
\end{equation}
The earlier observation that $a+b+c=1$ from Eq.~\eqref{eq:tbmtobtm} implies rather notable relationships among the sums of the entries in the rows and columns of $M_e$ itself, i.e., 
\begin{align}
&\sum_{i=1}^3(M_e)_{i1}=\sum_{i=1}^3(M_e)_{1i}=\left| m_{\tau }\right|^2(b^2+bc+ab)=b\left| m_{\tau }\right|^2,\\
&\sum_{i=1}^3(M_e)_{i2}=\sum_{i=1}^3(M_e)_{2i}=\left| m_{\tau }\right|^2(bc+c^2+ac)=c\left| m_{\tau }\right|^2,\\
&\sum_{i=1}^3(M_e)_{i3}=\sum_{i=1}^3(M_e)_{3i}=\left| m_{\tau }\right|^2(ab+ac+a^2)=a\left| m_{\tau }\right|^2. \label{eq:rowsums}
\end{align}
These are mildly interesting because the above three sums can be parameterized by $a$, $b$, and $c$.  Now that we have analyzed the effect of $U_e$ on the charged lepton mass matrix, we turn next to analyze the flavor symmetries in this framework.

\subsection{Implications for Lepton Flavor Symmetries \label{subsec:connectionnus}}

The neutrino mixing matrix being of TBM form implies that the non-trivial diagonal $Z_2\times Z_2$ elements $G_{i}^{\text{Diag}}=U_{\nu}^{\dagger}G_ iU_{\nu}$ ($i=1,2,3$) are also transformed to this basis\cite{bottomup}. In particular, the generators for the tribimaximal $Z_{2}\times Z_{2}$ group can be identified as
\begin{equation}\label{eq:nondiagSUdef}
U=G_{3}^{\text{TBM}}=
\left(\begin{array}{ccc}
-1 & 0  & 0 \\
 0 & 0  & -1 \\
 0 & -1 & 0
\end{array}\right),~~~~~
S=G_{2}^{\text{TBM}}=\frac{1}{3}
\left(\begin{array}{ccc}
-1 &  2 & 2 \\
 2 & -1 & 2 \\
 2 & 2  & -1
\end{array}\right).
\end{equation}
For non-degenerate charged lepton masses, we assume that the charged lepton symmetries must take one of the following minimal forms in the diagonal charged lepton basis:
\begin{equation}\label{eq:Telem}
\begin{aligned}
&T_1^\text{Diag}=\text{Diag}(1,\omega,\omega^2),~T_2^\text{Diag}=\text{Diag}(\omega^2, 1,\omega),~T_3^\text{Diag}=\text{Diag}(\omega,\omega^2,1),\\
&(T_1^\text{Diag})^2=\text{Diag}(1,\omega^2,\omega),~(T_2^\text{Diag})^2=\text{Diag}(\omega, 1,\omega^2),~(T_3^\text{Diag})^2=\text{Diag}(\omega^2,\omega,1),
\end{aligned}
\end{equation}
where $\omega=e^\frac{2\pi i}{3}$. 

Using the definitions for the $T_i$ in Eq.~\eqref{eq:Telem}, it easy easy to see that
\begin{equation}\label{eq:Tirel}
\omega^2 T_1^\text{Diag}=T_2^\text{Diag}=\omega T_3^\text{Diag} ~~~~~~\text{and}~~~~~~~\omega (T_1^\text{Diag})^2=(T_2^\text{Diag})^2=\omega^2 (T_3^\text{Diag})^2.
\end{equation}
Therefore, it is possible to transform from one $T_i^\text{Diag}$ to another (or from one $(T_i^\text{Diag})^2$ to another) by multiplying by a power of $\omega$.

Now that the forms of $T^\text{Diag}_{1,2,3}$ and $(T_{1,2,3}^\text{Diag})^2$ have been defined in Eq.~\eqref{eq:Telem}, it is possible to transform them to their undiagonalized form via the unitary transformation $U_{e}$ from Eq.~\eqref{eq:Utbtobtabc} such that
\begin{equation}
\label{eq:Ti}
T_{i} = U_{e}T^{\text{Diag}}_i U_{e}^{\dag}, 
\end{equation}
with $i=1,2,3$.  Notice that it is only necessary to undiagonalize one of the $T_i^\text{Diag}$ because it is possible to obtain the others just by multiplying by powers of $\omega$, cf.~Eq.~\eqref{eq:Tirel}. The undiagonalized $T_i^2$ can be derived just by squaring the above relation.
Noticeably, in addition to satisfying $T_{i}^{3}=1$ and leaving the charged lepton mass matrix in Eq.~\eqref{eq:Mediagonalize} invariant, i.e.,  $T_{i}^{\dag}M_{e}T_{i}=M_{e}$ for $i=1,2,3$,  if this relationship is satisfied, so is $(T_{i}^{\dag})^2M_{e}T_{i}^2=M_{e}$ for $i=1,2,3$, making $T_i^2$ a symmetry of $M_e$ as well.  Now that we have the forms for the $T_i$, it is necessary to check their relationships with $S$ and $U$, cf.~Eq.~\eqref{eq:nondiagSUdef}, in order to see if they can all close to form a group.

A straightforward calculation reveals that $S$, $U$, and \textit{all} $T_{i}$ obey
\begin{equation}\label{eq:rulesZeroDelta}
(UT_{i}^{-1}UT_{i})^{3}=1,~~~~~(SU)^{2}=1,~~~~~(ST_i)^{3}=1.
\end{equation}
This is because the $S$ and $U$ commute by definition, and the rules involving $T_i$ all involve powers which are a multiple of $3$, i.e., the order of $\omega$.  Surprisingly, $T_i\rightarrow T_i^2$ in the above relations also generates rules which are satisfied.  Yet it is found that \textit{only} $T_2$ and correspondingly $T_2^2$ satisfy
\begin{equation}\label{eq:rulesZeroDelta2}
(UT_{2})^{8}=(UT_{2}^2)^{8}=1.
\end{equation}
because $3$ does not divide $8$.  Therefore, transforming to $T_{1,3}$ or $T_{1,3}^2$ will involve additional powers of $\omega$ which will not vanish in the products.

The matrices $T_2$ and $T_2^2$ also satisfy another interesting relationship that $T_{1,3}$ do not, i.e.,
\begin{align}\label{eq:SfromUT}
S=U(UT_2)^4U(UT_2)^4=U(UT_2^2)^4U(UT_2^2)^4
=\frac{1}{3}
\left(\begin{array}{ccc}
	-1 &  2 & 2 \\
	2 & -1 & 2 \\
	2 & 2  & -1
\end{array}\right),
\end{align}
where $S$ and $U$ are as defined in Eq.~\eqref{eq:nondiagSUdef}.  Therefore, $S$ is not a generator for this group involving $U$ and $T_2$ because it can be written in terms of them.   Finally, since $T_2\leftrightarrow T_2^2$ does not change the relations given in Eqs.~\eqref{eq:rulesZeroDelta}-\eqref{eq:rulesZeroDelta2} nor Eq.~\eqref{eq:SfromUT}, without loss of generality define
\begin{equation}\label{eq:Tdef}
T= T_2.
\end{equation}
Notice that Eqs.~\eqref{eq:rulesZeroDelta}-\eqref{eq:rulesZeroDelta2} suggest that this group is $\Delta(96)$\cite{delta96pres}.\footnote{The relationship in Eq.~\eqref{eq:SfromUT} has already been discussed in $\Delta(96)$ in Ref.~\cite{delta96pres}.  Yet this relationship will prove to be a useful tool in the next section when CP violation is added.} However to be absolutely sure that the group is identified correctly, a scan over random products of up to length 8 of $T$, $S$, and $U$ of  $\mathcal{O}(10^4)$ iterations has been performed.  This scan reveals a group of 96 distinct elements, the number of elements in $\Delta(96)$. Additionally, the orders and traces/characters of the 96 elements have been checked.  They are all consistent with $\Delta(96)$. Therefore, we conclude this group must be $\Delta(96)$. It is important that the reader distinguishes between starting with $\Delta(96)$ to predict $U^{\text{BTM}}$ (as already done in the literature) and our approach, i.e., starting with charged lepton and neutrino mixing matrices themselves, identifying their corresponding symmetry generators, and deriving the algebraic relations among them to arrive at $\Delta(96)$. However, this analysis was done without considering CP violation.  In the next section, a CP-violating phase will be added to bitrimaximal mixing in order to determine if and how this changes the flavor symmetry group.

%%%%%%%%%%%%%%%%%%%%%%%%%%%%%%%%%%%%%%%%%%%%%%%%%%
\section{Imposing CP Violation with BTM Mixing\label{sec:btmphase}}

In the previous section, it was shown that it is possible to start with tribimaximal mixing in the neutrino sector and to obtain bitrimaximal mixing through a non-trivial charged lepton mixing matrix $U_e$ (see Eqs.~\eqref{eq:tbmtobtm}-\eqref{eq:Utbtobt}). Thus, it is possible to start with a vanishing reactor mixing angle in the neutrino mixing matrix  and through a non-trivial charged lepton mixing matrix generate a larger reactor angle more consistent with the observed data. However, a limitation of the previous approach is that the resulting bitrimaximal mixing matrix does not incorporate non-trivial CP violation.  Therefore, in this section CP violation will be incorporated in the previous framework by manually adding a CP-violating phase, $\delta$, to $U^{\text{BTM}}$ by assuming that $\delta$ has its origin in a discrete flavor symmetry group in the UV which breaks down to the residual discrete group at low energies. Afterwards, the results of this addition on the flavor symmetry group will be explored.   However, before proceeding with this analysis, it is important to mention that while not committing to a particular model,  a vanishing $\delta$ at a high-energy scale can generate a non-vanishing $\delta$ at low-energy scales through renormalization group effects.\footnote{An example of such an effect is described in Ref.~\cite{runningAntusch} in the context of the Standard Model and the large $\tan \beta$ regime of the Minimal Supersymmetric Standard Model, where the Majorana phases drive the running of $\delta$ at leading order.}
%\footnote{Such analysis is performed in the basis of diagonal charged leptons though, as opposed to ours, see Eq.~\eqref{eq:tbmtobtm}.}.
In order to incorporate the CP-violating phase $\delta$ with BTM mixing angle predictions, the PDG convention for the MNSP matrix is assumed\cite{pdg} while using the BTM angle predictions.  Thus, the now $\delta$-dependent BTM mixing, $U_{e}^{\dag}(\delta)U^{\text{TBM}}$, extends Eq.~(\ref{eq:Utbtobt}) to the case of non-zero $\delta$ as
\begin{equation}
U_{e}\to U_{e}(\delta)=U^{\text{TBM}}[U^{\text{BTM}}(\delta)]^{\dag},
 \label{eq:complexUtbtobt}
\end{equation}
where in a slight abuse of notation the BTM label has been kept for the corresponding MNSP matrix. It is worth noting that $U^{\text{BTM}}(\delta)$ has already appeared in Equation (40) of Ref. \cite{presneut} as
\begin{align}
&U^{\text{BTM}}(\delta)=
&\left(
\begin{array}{ccc}
 \frac{1}{6} \left(3+\sqrt{3}\right) & \frac{1}{\sqrt{3}} & \frac{1}{6} \left(3-\sqrt{3}\right) e^{-i \delta } \\
 \frac{3 \left(\sqrt{3}-1\right)+\left(2 \sqrt{3}-3\right) e^{i \delta }}{6 \sqrt{3}-15} & \frac{-3+\left(9-5 \sqrt{3}\right) e^{i \delta }}{6 \sqrt{3}-15}
   & \frac{1}{\sqrt{3}} \\
 \frac{1}{78} \left(12 \left(\sqrt{3}-4\right)+\left(9+\sqrt{3}\right) e^{i \delta }\right) & \frac{\left(3-2 \sqrt{3}\right) e^{i \delta }+3-3 \sqrt{3}}{6
   \sqrt{3}-15} & -\frac{1}{6} \left(3+\sqrt{3}\right) \\
\end{array}\right), \label{eq:btmdelta}
\end{align}
under a different scenario where charged leptons were assumed to be diagonal.\footnote{For an equivalent form of $U^{\text{BTM}}(\delta)$, in terms of the parameters $a$ and $b$ defined in Eqs.~\eqref{eq:abcdef}-\eqref{eq:Utbtobtabc}, see Appendix \ref{sec:appDeltaMatrices}.} 

The already cumbersome $U^{\text{BTM}}(\delta)$ in Eq.~\eqref{eq:btmdelta} guarantees a similarly complicated form for $U_{e}(\delta)$, whose full entries are listed in Appendix \ref{sec:appDeltaMatrices} because they are not particularly illuminating. The inclusion of a CP-violating phase affects the $T_{i}$ as well by promoting Eq.~\eqref{eq:Ti} to
\begin{equation}\label{eq:Tedelta}
T_{i}\to T_{i}(\delta)\equiv U_{e}(\delta)T^{\text{Diag}}_i[U_{e}(\delta)]^{\dag}. 
\end{equation}
In the spirit of Eq.~(\ref{eq:Tdef}), the $T_{2}$ permutation is chosen and we will simply refer to $T_{2}(\delta)$ as $T(\delta)$. Observe that because the elements in the neutrino sector, $S=G_{2}^{\text{TBM}}$ and $U=G_3^{\text{TBM}}$,  contain no CP violation by construction due to the fact that the tribimaximal mixing pattern has zero CP violation, the $T$ element must acquire the CP phase through the $U_{e}(\delta)$ transformation, see Eq.~(\ref{eq:tbmtobtm}).

A tedious check shows that not only is $T(\delta)$ is an order-3 element, but it also satisfies the following algebraic rules with $U$ as defined in Eq.~(\ref{eq:nondiagSUdef}), i.e.,
\begin{equation}\label{eq:rulesNonzeroDelta}
[UT^{-1}(\delta)UT(\delta)]^{3}=1,~~~~~[UT(\delta)]^{8}=1.
\end{equation}
Remarkably, further algebraic relations can be obtained. To see this, consider the following long product formed from the $U$ and $T(\delta)$ generators,
\begin{equation}
\mathcal{S}(\delta)\equiv U\bigl(UT(\delta)\bigr)^{4}U\bigl(UT(\delta)\bigr)^{4} \label{eq:long},
\end{equation}
which is denoted by $\mathcal{S}$ in analogy to Eq.~(\ref{eq:SfromUT}). One can demonstrate that $\mathcal{S}$ is an order-2 element and that
\begin{equation}\label{eq:rulesNonzeroDelta2}
\bigl[ \mathcal{S}(\delta) T(\delta)\bigr]^{3}=1,~~~~~[\mathcal{S}(\delta)U]^{2}=1.
\end{equation}
As a result, the $\delta=0$ algebraic relations in Eq.~(\ref{eq:rulesZeroDelta}) and Eq.~(\ref{eq:rulesZeroDelta2}) generalize to the $\delta\neq0$ relations in Eq.~(\ref{eq:rulesNonzeroDelta}) and Eq.~(\ref{eq:rulesNonzeroDelta2}), unmistakably exhibiting the preservation of $\Delta(96)$ as the underlying flavor symmetry group associated with this mixing form.  Additionally, it has been checked that $T_2(\delta)\leftrightarrow T_2^2(\delta)$ leaves the above relations unchanged as in the $\delta=0$ case.    However, it must be pointed out that while $\mathcal{S}(\delta)\to S=G_{2}^{\text{TBM}}$ when $\delta \to 0$, the first relation in Eq.~\eqref{eq:rulesNonzeroDelta2} does not hold if $\mathcal{S}(\delta)$ is replaced with $S=G_{2}^{\text{TBM}}$ when the phase does not vanish.\footnote{A similar result appears in Ref.~\cite{presneut} in the context of a strictly diagonal charged lepton basis where BTM mixing originates solely in the neutrino sector.  Here it was also observed that $U(UT)^{4}U(UT)^{4}\neq G_{2}^{\text{BTM}}(\delta)$, where $U=G_{3}^{\text{BTM}}(\delta)$ and $T$ is diagonal.}  Still, it is clear that $\mathcal{S}(\delta)$ is a group element whose order follows from the presentation rules of $\Delta(96)$.

In summary, the relations in Eq.~\eqref{eq:rulesNonzeroDelta} are satisfied for $U=G_3^{\text{TBM}}$ and $T(\delta)=T_2(\delta)$ for every value of $\delta$. Additionally, notice that $(UT^{-1}(\delta)UT(\delta))^3=1$ for any value of $\delta$ (see Eq.~\eqref{eq:Tdef} and surrounding text). Furthermore, $U$ and $T(\delta)$ can be used to define a $\delta$-dependent group element $\mathcal{S}(\delta)$ as was done in Eq.~\eqref{eq:long}.  As expected this product of generators is order 2.  Lastly, this element $\mathcal{S}(\delta)$ can be used to verify two more $\delta$-dependent relations given in Eq.~\eqref{eq:rulesNonzeroDelta2}, cf.~Eq.~\eqref{eq:rulesZeroDelta} for the two analogous relations with $\delta=0$.  Even though the CP-violating phase $\delta$ was imposed by hand in the bitrimaximal MNSP matrix given in Eq.~\eqref{eq:btmdelta}, this matrix is identical to Eq.~(40) of Ref.~\cite{presneut}. There we argued how $\delta$, despite appearing in the Jarlskog invariant of $U_{\nu}^{BTM}(\delta)$,
\begin{equation}
J=\dfrac{1}{234}(9+\sqrt{3})\sin{(\delta)}, \label{eq:jarlskog}
\end{equation}
cannot be granted a physical interpretation when the flavor symmetry is $\Delta(96)$ because this group admits a basis where all Clebsch-Gordan coefficients are strictly real. This supersedes the low-energy result $J\neq0$ and rules out a CP-violating phase of group theoretical origin. In the previous work, we started already with the presentation rules for $\Delta(96)$ and found that a spurious (unphysical) phase was unfixed by such rules. In the current work, the unphysical nature of $\delta$ can also be shown because any value that $\delta$ takes yields the same relations among the generators.  This implies the $\delta$ plays no role in defining this group. Therefore, changing the group's basis cannot select a specific value for the CP-violating phase $\delta$, and hence this phase cannot be physical.  Yet, in both approaches, bitrimaximal mixing is found to be obtained from $\Delta(96)$ and cannot accommodate the addition of a non-trivial  CP-violating phase\footnote{See Ref.~\cite{bottomupext} for a discussion of the role of unphysical phases on flavor symmetry group generators.} (see Refs.~\cite{presneut} and \cite{nocp} for different confirmations of this result).

%%%%%%%%%%%%%%%%%%%%%%%%%%%%%%%%%%%%%%%%%%%%%%%%%%%%%%%%%%%%%%%%%%%%%%%%%%%%%%%%%%%%%%%%%%%%%%%%%%%%%%%%%%%%%%%%%%%%%%%%%%%%%%%%%%%%%%%%%%%%%%%%%%%%%%

\section{Conclusions\label{sec:conclusions}}

%%%%%%%%%%%%%%%%%%%%%%%%%%%%%%%%%%%%%%%%%%%%%%%%%%%%%%%%%%%%%%%%%%%%%%%%%%%%%%%%%%%%%%%%%%%%%%%%%%%%%%%%%%%%%%%%%%%%%%%%%%%%%%%%%%%%%%%%%%%%%%%%%%%%%%
A trimaximal middle column is a feature common to both bitrimaximal and tribimaximal mixing patterns.  Usually, these patterns are derived by considering a diagonal charged lepton sector and obtaining all of the mixing from the neutrino sector.  In the current work, this assumption was relaxed in order to explore the implications of starting with tribimaximal mixing in the neutrino sector and arriving at the bitrimaximal mixing MNSP form through the non-trivial charged lepton mixing matrix given in Eq.~\eqref{eq:Utbtobt}. To this end, we began by looking closely at trimaximality, the common feature between the BTM and TBM mixing patterns. Using the explicit representations for the $Z_2\times Z_2$ formalism developed in Ref.~\cite{presneut}, we then extracted a parametric dependence for the solar and atmospheric angles in terms of $\theta_{13}$ (see Fig.~\eqref{fig:mxplot}).

To delve deeper into the connection between both mixing patterns, we started with the group elements for tribimaximal mixing in the neutrino sector, see Eq.~\eqref{eq:nondiagSUdef}.\footnote{The freedom of working in a diagonal charged lepton basis for $U_{\text{MNSP}}$ is reviewed in Appendix \ref{sec:appFreedom}.} Then, we calculated the non-trivial unitary transformation, $U_e$, that must exist in the charged lepton sector which would render the MNSP matrix in its bitrimaximal form, c.f.~Eqs.~\eqref{eq:undiagMe}-\eqref{eq:Utbtobtabc}. The transformation $U_{e}$ and the corresponding undiagonalized charged lepton mixing matrix both display peculiar algebraic relations for the sums of its rows and columns, as indicated in Secs. \ref{subsec:connection} and \ref{subsec:implicationsCharged}. These relations deserve further study.

Moving on to the charged lepton sector, we assumed the smallest possible symmetry which can generate three non-degenerate charged lepton masses, i.e., $Z_3$, by positing all possible diagonal forms for this matrix, cf., Eq.~\eqref{eq:Telem}. We then undiagonalized these six diagonal matrices subject to Eq.~\eqref{eq:Ti} to yield a set of six $T_i$. We were then able to brute-force calculate a set of presentation rules between the $T_{1,2,3}$, $S$, and $U$, cf.~Eq.~\eqref{eq:rulesZeroDelta}, by scanning over powers of products of these matrices. When deriving the relationship that exists between $U$ and $T_{1,2,3}$ in Eq.~\eqref{eq:rulesZeroDelta2}, it was found that such a relationship could only be satisfied if $T=T_2$ or $T=T_2^2$ and their products $(UT_2)$ or  $(UT_2^2)$ are of order 8. This further restricted the discussion of the charged lepton symmetry elements to only $T_2$ and $T_2^2$. Since the relations/presentation rules did not change under the interchange of $T_2$ and $T_2^2$, we further narrowed the discussion to only consider $T_2$ in Eqs.~\eqref{eq:rulesZeroDelta}-\eqref{eq:SfromUT}.

Next motivated by the discussion at the end of our previous work \cite{presneut}, we analyzed the incorporation of a CP-violating phase in our results for $U_{e}$. The starting point consisted of manually placing the CP-violating phase $\delta$ in the $U^{\text{BTM}}$ in accordance with the MNSP parametrization described in Ref.~\cite{bottomup}.  We then looked at the consequences for the now phase-dependent $U_{e}(\delta)$ and $T(\delta)$. Amusingly, we found that not only $U$ and $T(\delta)$ satisfy the $\delta$-dependent analogues of the presentation rules in Eqs.~\eqref{eq:rulesZeroDelta}-\eqref{eq:rulesZeroDelta2}, but the product $U[UT(\delta)]^{4}U[UT(\delta)]^{4}$ also satisfies the same relations with $U$ and $T(\delta)$ as in the last two relations of Eq.~\eqref{eq:rulesZeroDelta}. The unavoidable conclusion is that exactly the same flavor group is preserved under the introduction of the CP-violating phase. Unfortunately, the presentation rules in Eq.~\eqref{eq:rulesNonzeroDelta} and the further relations in Eqs.~\eqref{eq:long}-\eqref{eq:rulesNonzeroDelta2} still hold (and hence the imply the same flavor symmetry group) regardless of the value of the phase, including $\delta=0$. This leads to the unavoidable conclusion that $\delta$ fails to be a physical phase under the set of assumptions taken, i.e., three non-degenerate, non-zero Majorana neutrinos, a residual $Z_{n}$ ($n\geq 3$) symmetry in the charged lepton sector, and a tribimaximal neutrino mixing matrix $U_{\nu}$,\footnote{Actually Appendix \ref{sec:appFreedom} shows it is possible to relax this assumption.} leading to a  bitrimaximal $U^{\text{MNSP}}$. The paradigm developed in this work can be immediately applied to other choices of charged lepton and neutrino mixing patterns. Furthermore, the results of this study can be used to help pinpoint the underlying flavor symmetry group and how it may change into another symmetry group upon the manual introduction of a CP-violating phase.  In both cases, further research into other lepton mixing choices which may be exploited to obtain a prediction for $\delta$ is left to future work.

%%%%%%%%%%%%%%%%%%%%%%%%%%%%%%%%%%%%%%%%%%%%%%%%%%%%%%%%%%%%%%%%%%%%%%% 

\section*{Acknowledgments}

%%%%%%%%%%%%%%%%%%%%%%%%%%%%%%%%%%%%%%%%%%%%%%%%%%%%%%%%%%%%%%%%%%%%%%% 

We thank A.~Aranda, L.~Everett, and M.~J.~Pérez for helpful comments and suggestions.   C.A. would like to thank the Facultad de Ciencias for their hospitality in the early stage of this work.  A.J.S. and J.B. would like to acknowledge partial support from CONACYT project CB2017-2018/A1-S-39470 (M\'exico).

\appendix
\section{MNSP Basis Freedom and Flavor Symmetry Groups}
\label{sec:appFreedom}
Equation \eqref{eq:tbmtobtm} is a particular choice of a starting neutrino mixing matrix (TBM). One could start from an arbitrary neutrino-sector transformation $U_{\nu}$, and bring the MSNP to the bitrimaximal form through
\begin{equation}\label{eq:BTMagainTBM}
	U_{\text{MNSP}}\equiv U^{\text{BTM}}=U_{e}^{\dagger}U_{\nu}.
\end{equation}
The corresponding generators for the charged lepton sector and the neutrino sector are $T=T_2$, $S=G_{2},$ and $U=G_{3}$. It has been shown in Ref.~\cite{bottomup} and reviewed in Section \ref{subsec:connection} that this set of non-diagonal generators are obtained from their diagonal counterparts through $S=U_{\nu}G_{2}^{\text{Diag}}U_{\nu}^{\dagger}$, $U=U_{\nu}G_{3}^{\text{Diag}}U_{\nu}^{\dagger}$, and $T= U_{e} T^{\text{Diag}}_2 U_{e}^{\dagger}$,
where $G_{2}^{\text{Diag}}=\text{Diag}(-1,1,-1)$, $G_{3}^{\text{Diag}}=\text{Diag}(-1,-1,1)$, 
and $T_{2}^{\text{Diag}}$ defined in Eq.~\eqref{eq:Telem}. $S$, $T$, and $U$ satisfy the presentation rules
\begin{equation}\label{eq:presenRulesG3T}
	\langle U,T \mid
	U^{2}=T^{3}=(UT)^{8}=(UT^{-1}UT)^{3}=1
	\rangle,
\end{equation}
and the relation
\begin{align}
S=U(UT)^{4}U(UT)^{4}.
\end{align}
However, the presentation also holds when using the charged lepton generator in its diagonalized form $T_2^{\text{Diag}}$, by redefining the two $Z_2\times Z_2$ group generators in the neutrino sector as follows:
\begin{align*}
S'&\equiv U_{e}^{\dagger} S U_{e}=U_{e}^{\dagger} (U_{\nu} G_{2}^{\text{Diag}}  U_{\nu}^{\dagger} ) U_{e}= U^{\text{BTM}} G_{2}^{\text{Diag}}U^{\text{BTM}\dagger}, \\
U'&\equiv U_{e}^{\dagger} U U_{e}=U_{e}^{\dagger} (U_{\nu} G_{3}^{\text{Diag}}  U_{\nu}^{\dagger} ) U_{e}= U^{\text{BTM}} G_{3}^{\text{Diag}}U^{\text{BTM}\dagger}.
\end{align*}
With this said, it becomes straightforward to see that
\begin{equation*}
U^{2}=U_{\nu} G_{3}^{\text{Diag}} U_{\nu}^{\dagger} U_{\nu} G_{3}^{\text{Diag}} U_{\nu}^{\dagger}=U_{\nu}(G_{3}^{\text{Diag}})^{2} U_{\nu}^{\dagger}
\end{equation*}
implies that $(U')^{2}=1$ iff $U^{2}=1$. Similarly, we notice the trivial case
\begin{equation*}
(T)^{3}=U_{e} T^{\text{Diag}} U_{e}^{\dagger} U_{e} T^{\text{Diag}} U_{e}^{\dagger} U_{e} T^{\text{Diag}} U_{e}^{\dagger}=U_{e}(T^{\text{Diag}})^{3} U_{e}^{\dagger}.
\end{equation*}
Moreover, by replacing Eqs.~\eqref{eq:BTMagainTBM}-\eqref{eq:presenRulesG3T} in the remaining presentation rules, we show their direct relation to $U'$ and $T^{\text{Diag}}$
\begin{align*}
	(U T)^{8}
	&= \left( U_{\nu} G_{3}^{\text{Diag}}  U_{\nu}^{\dagger} U_{e} T^{\text{Diag}} U_{e}^{\dagger} \right)^{8}\\
	&=\left(U_{\nu} G_{3}^{\text{Diag}} (U^{\text{BTM}})^{\dagger} T^{\text{Diag}} U_{e}^{\dagger}\right)^{8}\\
	&=\left(U_{\nu} G_{3}^{\text{Diag}}\left(U^{\text{BTM}}\right)^{\dagger} T^{\text{Diag}} U_{e}^{\dagger} U_{\nu} G_{3}^{\text{Diag}} (U^{\text{BTM}})^{\dagger} T^{\text{Diag}} U_{e}^{\dagger}\right)^{4}\\
	&=\left(U_{\nu}G_{3}^{\text{Diag}} (U^{\text{BTM}})^{\dagger} T^{\text{Diag}} U^{\text{BTM}} G_{3}^{\text{Diag}} (U^{\text{BTM}})^{\dagger} T^{\text{Diag}} U_{e}^{\dagger} \right)^{4}\\
	&=\left(U_{e} U' T^{\text{Diag}}  U' T^{\text{Diag}} U_{e}^{\dagger}\right)^{4}\\
	&=\left(U_{e} (U' T^{\text{Diag}})^{2} U_{e}^{\dagger}\right)^{4}\\
	&=U_{e} (U' T^{\text{Diag}})^{2}  U_{e}^{\dagger} U_{e}(U' T^{\text{Diag}})^{2}U_{e}^{\dagger} U_{e}(U' T^{\text{Diag}})^{2}  U_{e}^{\dagger} U_{e} (U' T^{\text{Diag}})^{2} U_{e}^{\dagger}\\
	&= U_{e} (U' T^{\text{Diag}})^{8} U_{e}^{\dagger},
\end{align*}
\begin{align*}
	(UT^{-1}UT)^{3}
	&=(UT^{2}UT)^{3}\\
	&=\left( U_{\nu} G_{3}^{\text{Diag}} U_{\nu}^{\dagger} U_{e} T^{\text{Diag}} U_{e}^{\dagger} U_{e} T^{\text{Diag}} U_{e}^{\dagger} U_{\nu} G_{3}^{\text{Diag}} U_{\nu}^{\dagger}  U_{e} T^{\text{Diag}} U_{e}^{\dagger} \right)^{3}\\
	& =\left(U_{\nu} G_{3}^{\text{Diag}} (U^{\text{BTM}})^{\dagger} (T^{\text{Diag}})^{2}  U^{\text{BTM}} G_{3}^{\text{Diag}} (U^{\text{BTM}})^{\dagger}  T^{\text{Diag}} U_{e}^{\dagger}
	\right)^{3}\\
	& =\left( U_{e} U' (T^{\text{Diag}})^{2} U' T^{\text{Diag}} U_{e}^{\dagger}
	\right)^{3}\\
	& = U_{e} \left(U' (T^{\text{Diag}})^{2} U' T^{\text{Diag}} \right)^3 U_{e}^{\dagger}\\
	& = U_{e} \left(U' (T^{\text{Diag}})^{-1} U' T^{\text{Diag}} \right)^3 U_{e}^{\dagger}.
\end{align*} 
Therefore,  $(UT)^{8}=1$ iff $(U' T^{\text{Diag}})^{8}=1$, and $(UT^{-1}UT)^{3}=1$ iff $U_{e} \left(U' (T^{\text{Diag}})^{-1} U' T^{\text{Diag}} \right)^3 U_{e}^{\dagger}=1$. It can be proven that $(UT)^{4}=U_e (U' T^{\text{Diag}})^{4} U_{e}^{\dagger}$.  Thus,
\begin{align*}
	S
	&=U(UT)^{4}U(UT)^{4} \\
	&=UU_{e} (U' T^{\text{Diag}})^{4} U_{e}^{\dagger} U U_{e} (U' T^{\text{Diag}})^{4} U_{e}^{\dagger}\\
	&=U_{\nu} G_{3}^{\text{Diag}} U_{\nu}^{\dagger}
	U_{e} (U' T^{\text{Diag}})^{4} U_{e}^{\dagger}
	U^{\nu} U^{\text{Diag}} U_{\nu}^{\dagger}
	U_{e} (U' T^{\text{Diag}})^{4} U_{e}^{\dagger}\\
	&=U_{\nu} G_{3}^{\text{Diag}} (U^{\text{BTM}})^{\dagger} (U' T^{\text{Diag}})^{4}
	U^{\text{BTM}} G_{3}^{\text{Diag}} (U^{\text{BTM}})^{\dagger}(U' T^{\text{Diag}})^{4} U_{e}^{\dagger}\\  
	&=U_{e} \left( U'(U' T^{\text{Diag}})^{4} U' (U' T^{\text{Diag}})^{4} \right) U_{e}^{\dagger}\\   
	&=U_{e} S' U_{e}^{\dagger}.    
\end{align*}
Hence, choosing between $T$ or $T^{\text{Diag}}$ is irrelevant for deriving the resulting flavor symmetry group in this paradigm, as long as everything is transformed accordingly. Ergo, this basis choice does not affect the flavor symmetry group generated by $S$, $T$, and $U$.  Furthermore because the starting point for $U_{\nu}$ was arbitrary, it is very clear from the above proof that it is the resulting MNSP matrix  $U_{\text{MNSP}}=U_e^{\dagger}U_{\nu}$ that really defines the flavor symmetry and not merely the basis-dependent forms for $U_e$ and $U_{\nu}$.

%%%%%%%%%%%%%%%%%%%%%%%%%%%%%%%%%%%%%%%%%%%%%%%%%%%%%%%%%%%%%%%%%%%%%%% 

\section{$\mathbf{U_e(\delta)}$ and an Alternative Expression for $\mathbf{U^{BTM}(\delta)}$\label{sec:appDeltaMatrices}}

%%%%%%%%%%%%%%%%%%%%%%%%%%%%%%%%%%%%%%%%%%%%%%%%%%%%%%%%%%%%%%%%%%%%%%% 

The complete expressions for entries of the $U_{e}(\delta)$ defined in Eq.~\eqref{eq:complexUtbtobt} have the forms
\begin{align}
	U_{e}(\delta)_{11} &= \dfrac{1}{6}\bigl(2+\sqrt{2}+\sqrt{6}\bigr), \\
	U_{e}(\delta)_{12} &= \dfrac{1}{3(-6+5\sqrt{3})}\bigl( 3+3\sqrt{2}-3\sqrt{6}+(-9+3\sqrt{2}+5\sqrt{3}-2\sqrt{6})e^{-i\delta} \bigr), \\
	U_{e}(\delta)_{13} &= \dfrac{1}{36-30\sqrt{3}}\bigl( -6(-1-4\sqrt{2}+\sqrt{3}+2\sqrt{6})+(6-3\sqrt{2}-4\sqrt{3}+\sqrt{6})e^{-i\delta} \bigr), \\
	U_{e}(\delta)_{21} &= \dfrac{1}{36}\bigl( -3(-4+\sqrt{2}+\sqrt{6})-3\sqrt{2}(-3+\sqrt{3})e^{i\delta} \bigr), \\
	U_{e}(\delta)_{22} &= \dfrac{1}{6(-5+2\sqrt{3})}\bigl( \sqrt{6}(-4-\sqrt{2}+\sqrt{3})+(-10-2\sqrt{2}+6\sqrt{3}+\sqrt{6})e^{-i\delta} \bigr), \\
	U_{e}(\delta)_{23} &= \dfrac{1}{-60+24\sqrt{3}}\bigl( -12+21\sqrt{2}+4\sqrt{3}-9\sqrt{6}+(-8-\sqrt{2}+4\sqrt{3}+\sqrt{6})e^{-i\delta} \bigr), \\
	U_{e}(\delta)_{31} &= \dfrac{1}{36}\bigl( -3(-4+\sqrt{2}+\sqrt{6})+3\sqrt{2}(-3+\sqrt{3})e^{i\delta} \bigr), \\
	U_{e}(\delta)_{32} &= \dfrac{1}{6(-5+2\sqrt{3})}\bigl( -9\sqrt{2}-2\sqrt{3}+6\sqrt{6}+(-10-2\sqrt{2}+6\sqrt{3}+\sqrt{6})e^{-i\delta} \bigr), \\
	U_{e}(\delta)_{33} &= \dfrac{1}{-60+24\sqrt{3}}\bigl( -12+3\sqrt{2}+4\sqrt{3}-7\sqrt{6}+(-8-\sqrt{2}+4\sqrt{3}+\sqrt{6})e^{-i\delta} \bigr).
\end{align}
Finally, observe that an equivalent form for $U^{\text{BTM}}(\delta)$, cf.~Eq.~\eqref{eq:btmdelta}, which has no radicals in the denominators can be written as
\begin{align}\nonumber
&U^{\text{BTM}}(\delta)=\\\nonumber
&\left(
\begin{array}{ccc}
 \frac{1}{6} \left(3+\sqrt{3}\right) & \frac{1}{\sqrt{3}} & -\frac{1}{6} \left(\sqrt{3}-3\right) e^{-i \delta } \\
 \tfrac{1}{13}\bigl( -(1+3\sqrt{3})+\tfrac{\sqrt{3}}{3}(-4+\sqrt{3})e^{i\delta} \bigr) & \tfrac{1}{13}\bigl( (5+2\sqrt{3})+\tfrac{1}{3}(-15+7\sqrt{3})e^{i\delta} \bigr)
   & \frac{1}{\sqrt{3}} \\
 \tfrac{1}{13}\bigl( 2(-4+\sqrt{3})+\tfrac{\sqrt{3}}{6}(1+3\sqrt{3})e^{i\delta} \bigr) & \tfrac{1}{13}\bigl( (1+3\sqrt{3})-\tfrac{\sqrt{3}}{3}(-4+\sqrt{3})e^{i\delta} \bigr) & \frac{1}{6} \left(-3-\sqrt{3}\right) \\
\end{array}\right).
\end{align}
This form of this matrix can be expressed in terms of the parameters $a$ and $b$ from Eqs.~\eqref{eq:abcdef}-\eqref{eq:Utbtobtabc} due to the relations
\begin{align}
\left(1+3\sqrt{3}\right) &= 1+\dfrac{a-b}{ab}, \\
\left(-4+\sqrt{3}\right) &= -1-\dfrac{2}{3}\dfrac{b-c}{a b}, \\
\left(5+2\sqrt{3}\right) &= \left(1+\dfrac{1-3b}{\sqrt{2}}\right)\left(1-\dfrac{1}{a}+\dfrac{1}{b}\right), \\
\left(-15+7\sqrt{3}\right) &= \dfrac{8-19b-a(5+9b)}{3 ab}.
\end{align}

%%%%%%%%%%%%%%%%%%%%%%%%%%%%%%%%%%%%%%%%%%%%%%%%%%%%%%%%%%%%%%%%%%%%%%%%%

\end{document}